\documentclass[final,3p,times,twocolumn]{elsarticle}
\usepackage{graphicx}
\usepackage{amssymb}
\usepackage{amsmath}
\usepackage{hyperref}
\usepackage{gensymb}

\usepackage{amsmath}
\usepackage{makeidx}
 
 \journal{Physica D (accepted)}

\usepackage[usenames]{color}
\usepackage[normalem]{ulem}
\usepackage{gensymb}

\usepackage{hyperref}
\usepackage[T1]{fontenc} 
\begin{document}

\title{Expansion dynamics of a  cylindrical-shell-shaped strongly dipolar 
condensate}

\author[fac]{Luis E. Young-S.}
\ead{lyoung@unicartagena.edu.co}

 \address[fac]{Grupo de Modelado Computacional, Facultad de Ciencias Exactas y Naturales, Universidad de Cartagena, 
130015 Cartagena de Indias, Bolivar, Colombia}

 \author[int]{S. K. Adhikari}
\ead{sk.adhikari@unesp.br}

\address[int]{Instituto de F\'{\i}sica Te\'orica, UNESP - Universidade Estadual Paulista, 01.140-070 S\~ao Paulo, S\~ao Paulo, Brazil}

\begin{abstract}

A Bose-Einstein condensate (BEC) formed on a curved surface with a distinct topology has been a hot topic of intense research, in search of new phenomena in quantum physics as well as for its possible application in quantum computing.  In addition to  the study of a spherical-shell-shaped BEC, we studied the formation of a cylindrical-shell-shaped harmonically-trapped  dipolar BEC of  $^{164}$Dy atoms theoretically using an improved mean-field model  including a Lee-Huang-Yang-type interaction, meant to stop a collapse at high
atom density.  To test the robustness of the cylindrical-shell-shaped BEC, here  we study its expansion in the same model.  We find that as the harmonic trap in the $x$ and $y$ directions are removed, maintaining the axial trap,  the cylindrical-shell-shaped BEC expands in the $x$-$y$ plane  without deformation, maintaining its shell-shaped structure.   After an adequate radial expansion, the axial trap can be relaxed for a desired axial expansion of the cylindrical-shell-shaped BEC   allowing its observation.

\end{abstract}


\maketitle

\section{Introduction}

A Bose-Einstein condensate (BEC) of bosonic atoms  has been  fundamental in investigating  different quantum phenomena, which, although previously predicted and studied theoretically,   could  not be realized experimentally  otherwise.  For example, to mention a few, a rotating BEC has generated a very large vortex lattice  confirming its superfluidity \cite{vl};  quantum phase transition has been realized in a BEC on an optical lattice \cite{qpt1,qpt2}; {bright
matter-wave solitons} have been studied in a quasi-one-dimensional attractive BEC  \cite{qs1,qs2};  the effect of spin-orbit coupling of neutral atoms 
on a spinor BEC has been investigated \cite{so};  a quasi-two-dimensional (quasi-2D)  supersolid BEC has been studies experimentally \cite{ss1,ss2} .  
Most of these studies were performed in the three-dimensional (3D) Eucledean space with trivial topology. 
 Nevertheless, 
new physics is expected to  emerge on curved  surfaces with distinct topology, for example, on the surface of a hollow sphere \cite{BBL-space,EO,lab,lab2}  or a hollow cylinder \cite{hc}, the latter, although   topologically equivalent to a toroid or a ring, has a distinct 
geometrical shape.   Generation of vortices on a  curved surface is distinct from the same in a 3D space \cite{v1,v2,v3}.
 Unique superfluid properties may exist in a BEC in a toroidal 
trap with distinct topology  \cite{r1,r2,r3}. Moreover, 
 quantum states with distinct topology  \cite{top1,top2,top3,curved}  may have useful application in the field of quantum computing \cite{QC}.

In view of the above-mentioned interests in a BEC with a distinct topology, experimental activity has started  to create and study   a BEC formed on a spherical surface in the form of a spherical bubble.
 In the presence of gravity, even in a spherical-shell-shaped trap, this is not possible to create a spherical bubble-shaped BEC,
 as
gravity will bring down the atoms from the top of the bubble making a hole.  However,  partially circumventing this problem,
a hemispherical-shell-shaped BEC of $^{87}$Rb atoms   has been created in a laboratory on earth \cite{bs2}  by  a  gravity compensation mechanism.  As a first step  to create
a BEC   in a spherical-shell-shaped trap, 
a spherical bubble of ultracold $^{87}$Rb atoms 
 in  orbital microgravity \cite{BBL-space} in a  orbiting space station, without the effect of gravity,   has been realized  \cite{EO,lab,lab2} 
 following  a suggestion  \cite{sugg1,sugg2} and studies in Ref.  \cite{sugg3,sugg4,sugg5,TS}.  
 Following another theoretical suggestion  \cite{Ho,Pu} and related investigations \cite{bin1,bin2,bin3,gammal}, 
 a binary BEC of $^{23}$Na and $^{87}$Rb atoms have been observed in a laboratory on earth  \cite{expan3}, in the presence of gravity,  with the $^{23}$Na atoms forming a spherical bubble surrounding a solid spherical core of  $^{87}$Rb atoms. These experiments \cite{BBL-space,bs2,expan3} for the creation of a BEC with a distinct topology 
 in the form of a spherical bubble require a complex manipulation of external conditions or of the confining trap.
 
 Recently, we have demonstrated \cite{hc} the possibility of the formation of a cylindrical-shell-shaped BEC of strongly dipolar $^{164}$Dy atoms, with distinct topology,  in a  {\it harmonic} trap, not requiring a cylindrical-shell-shaped trap.   Although a cylindrical shell is topologically equivalent to a circular ring, it has a different geometric shape.  This may allow the possibility of the study of new physics on the surface of a cylinder, like the creation of vortex or vortex lattice.  In this paper we test the robustness of these 
 cylindrical-shell-shaped
 states by real-time propagation during an expansion for a long period of time. We demonstrate that the cylindrical-shell-shaped  state expands  without destroying the topology to a very large size, where it can be photographed and its presence be experimentally confirmed. For the detection of a shell-shaped   BEC,  its expansion dynamics is of utmost importance as emphasized in different studies in the context of a spherical-shell-shaped state \cite{expan3,expan2,expan4,expan1}.

  The theoretical investigation of this paper will be based on   an improved mean-field
 model including a Lee-Huang-Yang-type (LHY) interaction \cite{lhy}, appropriately modified for dipolar atoms \cite{qf1,qf3,qf2}.  In a mean-field Gross-Pitaevskii (GP) model with cubic nonlinearity, a strongly dipolar BEC  
 collapses  \cite{ex1,ex2},  and the inclusion of  a repulsive LHY   interaction \cite{qf1,qf2}
in theoretical investigations     stabilizes a strongly   dipolar
condensate   \cite{santos} to form  a droplet \cite{ex1,ex2,29}  in a strong harmonic  trap.  For a large number of strongly dipolar atoms, in a strong quasi-2D harmonic trap,  
a spatially-periodic  \cite{ss2}  triangular- \cite{ss2,ex1,ex2}, square- \cite{th4},   or  honeycomb-lattice \cite{th4} structure of droplets   or a  labyrinthine state  \cite{pfau} could be formed. 
 The present  cylindrical-shell-shaped $^{164}$Dy
 BEC is formed in a strong three-dimensional harmonic trap, and not a quasi-2D trap.
 As  the number of atoms is increased beyond  a critical value, the dipolar atoms are deposited on the outer surface of a hollow cylinder, for scattering length $a$ in the range 
 $85a_0  \gtrapprox a \gtrapprox80a_0$ \cite{hc}, where $a_0$ is the Bohr radius.  For the cylindrical-shell-shaped states to appear, 
  the  number of atoms $N$ should be in the range $250000 \gtrapprox N  \gtrapprox 150000$,
  the trap frequency $f_z$ along the polarization $z$ direction  in the range $ 250$ Hz  $\gtrapprox f_z \gtrapprox 150$ Hz and  those in the transverse $x$-$y$ plane are taken as $f_x, f_y \approx  0.75 f_z$ \cite{hc}.
 If the confining harmonic trap  in the $x$-$y$ plane is made slightly anisotropic ($f_x\ne f_y$),  the  cylindrical shell of the 
strongly dipolar BEC becomes elliptical in nature.  These strongly dipolar BECs have a large number of atoms, not trivial for experimental realization \cite{expt}, specially in the overall   attractive regime due to  large atom loss by three-body interaction.
A way  to optimize the observation  of strongly dipolar BECs with a large number of atoms 
($N> 2 \times  10^5$), appropriate for the study of the present  cylindrical-shell-shaped BECs, has recently appeared in the literature \cite{arch}.

 The expansion of a cylindrical-shell-shaped BEC of circular or elliptic section  can be considered 
 as a two-step process. First, we consider its quasi-free expansion 
 by setting the transverse trap frequencies to zero ($f_x=f_y=0$), but maintaining the frequency in the polarization direction $f_z$ unchanged, employing real-time propagation. 
We find that the initial symmetry of the  cylindrical-shell-shaped    BEC is  maintained during the quasi-free expansion. 
 In this case, 
 as the radius of the cylindrical shell increases during this expansion, the length of the cylinder (along the $z$ direction) reduces. 
The reduction of the root-mean-square (rms) size in the $z$ direction
is accompanied by a periodic oscillation, controlled by the frequency of the trap in the $z$ direction, in agreement  with a theoretical prediction \cite{stringari}.
After an adequate radial expansion, the axial $z$ frequency of the trap $f_z$ can be reduced slightly for 
an axial expansion of the system  along $z$ direction to a desirable    
size to be observed experimentally. We verified that the $z$ length increases without destroying the  cylindrical-shell-shaped structure of the BEC (not elaborated in this paper).

The expansion dynamics of the present  cylindrical-shell-shaped strongly dipolar BEC is very different from the free expansion of  a spherical-shell-shaped nondipolar BEC.  In the latter case, unlike in the present case, the BEC  expands inwards and outwards thus filling in the entire space destroying the shell-shaped structure very quickly \cite{expan2}.  The same may happen in the case of a freely expanding nondipolar ring-shaped BEC \cite{expan5}  or a freely expanding dipolar toroid-shaped BEC {\cite{mukerje} }
with the same topology as the present cylindrical-shell-shaped BEC. The ring-shaped BEC again expands both inwards and outwards closing  the hole in the ring entirely, thus destroying ring-shaped structure. This sets a limitation on the direct experimental observation of a spherical-shell-shaped nondipolar BEC.  
There have been suggestions about how to maintain the spherical-shell-shaped structure using a binary BEC \cite{Ho,Pu,expan3} or employing matter-wave lensing techniques \cite{expan4}.  
The strong  long-range dipolar  repulsion between  atoms located at diagonally opposite positions in the same $x$-$y$ plane  
in the present cylindrical-shell-shaped-state stops the inward expansion, thus stabilizing the shell structure.

In Sec. II we present the improved mean-field model including the LHY interaction and reduce it to a dimensionless form. In addition, we provide an expression of the dimensionless energy functional of the system (energy per atom.)
In Sec. III we present our numerical results for trap frequency $f_z=167$ Hz.  This axial trap frequency was recently used in the study of a strongly dipolar $^{164}$Dy BEC \cite{ss2,prl2d}.  In addition to presenting the profile of the hollow cylindrical shape of the quasi-freely expanding shell-shaped $^{164}$Dy BEC, we also present the evolution of energy and axial and radial sizes of the expanding BEC.   
In Sec. IV we present a brief summary of our findings.

\section{Improved mean-field model}

 We consider a dipolar BEC of $N$ $^{164}$Dy  atoms,  polarized along the $z$ direction, of mass $m$ each, and with atomic scattering length $a$. The magnetic dipolar interaction between two dipolar atoms, of magnetic moment $\mu$ each, located at  $\bf r \equiv \{x,y,z\}$ and $\bf r' \equiv \{x',y',z 
'\}$   is 
\begin{align} \label{dip-pot}
U_{\mathrm{dd}}({\bf R})=\frac{\mu_0 \mu^2 }{4\pi} \frac{1-3\cos^2 \theta}{|{\bf R}|^3}\equiv \frac{3\hbar^2 a_{\mathrm{dd}}}{m}  \frac{1-3\cos^2 \theta}{|{\bf R}|^3},
\end{align} 
where  $\theta$ is the angle made by the vector $\bf R \equiv r - r'$ with the polarization
$z$ direction,  $\mu_0$ is the permeability of vacuum, and the dipolar length $ a_{\mathrm{dd}} = \mu_0\mu^2m/(12\pi \hbar^2)$ is a measure of the strength of dipolar interaction, whereas the scattering length $a$ is a measure of the strength of contact interaction.

The statics and dynamics  of a  cylindrical-shell-shaped BEC \cite{hc}  is governed   by the following  improved  mean-field
Gross-Pitaevskii  equation with the inclusion of  the dipolar and the LHY interaction \cite{dipbec,blakie,dip,yuka}
\begin{align}\label{eq.GP3d}
 \mbox i \hbar \frac{\partial \psi({\bf r},t)}{\partial t} &=\
{\Big [}  -\frac{\hbar^2}{2m}\nabla^2 
+U({\bf r})
+ \frac{4\pi \hbar^2}{m}{a} N \vert \psi({\bf r},t) \vert^2 \nonumber\\
& +\frac{3\hbar^2}{m}a_{\mathrm{dd}}  N
\int U_{\mathrm{dd}}({\bf R})
\vert\psi({\mathbf r'},t)\vert^2 d{\mathbf r}'  \nonumber\\
& +\frac{\gamma_{\mathrm{LHY}}\hbar^2}{m}N^{3/2}
|\psi({\mathbf r},t)|^3
\Big] \psi({\bf r},t),   \\
\label{trap}
U({\bf r})&=\frac{1}{2}m(\omega_x^2x^2+\omega_y^2y^2+\omega_z ^2z^2) ,
\end{align}
where 
 $\omega_x (\equiv 2\pi f_x), \omega_y (\equiv 2\pi  f_y), \omega_z (\equiv 2\pi  f_z)$  are the angular frequencies of the harmonic trap (\ref{trap})
 along $x,y,z$ directions, respectively.
 The axis of the cylindrical shell is aligned along the polarization $z$ direction of dipolar atoms.
The wave function $\psi({\bf r},t)$ at time $t$ 
is  normalized as $\int \vert \psi({\bf r},t) \vert^2 d{\bf r}=1.$  The strength of LHY interaction 
  $\gamma_{\mathrm{LHY}}$ is given by \cite{qf1,qf3,qf2}
\begin{align}\label{qf}
\gamma_{\mathrm{LHY}}= \frac{128}{3}\sqrt{\pi a^5} Q_5(\varepsilon_{\mathrm{dd}}), \quad \varepsilon_{\mathrm{dd}}= \frac{ a_{\mathrm{dd}}}{a},
\end{align}
where  the auxiliary function $ Q_5(\varepsilon_{\mathrm{dd}})$ includes the correction to the LHY interaction \cite{lhy}
due to the dipolar interaction and is given by \cite{qf3}
{
 \begin{align}
  Q_5(\varepsilon_{\mathrm{dd}})= (1- \varepsilon_{\mathrm{dd}})^{5/2}  \,  _2F_1\left( -\frac{5}{2}, \frac{1}{2} ;
  \frac{3}{2}; \frac{3\varepsilon_{\mathrm{dd}}}{\varepsilon_{\mathrm{dd}}-1}  \right),
 \end{align}
 where $_2F_1$ is the hypergeometric function \cite{hyper}.  Using an integral representation of this function \cite{hyper}, 
 the auxiliary function $Q_5$ can be written as \cite{blakie} }
\begin{align}\label{exa1}
 Q_5(\varepsilon_{\mathrm{dd}})&=\ \int_0^1 dx(1-\varepsilon_{\mathrm{dd}}+3x^2\varepsilon_{\mathrm{dd}})^{5/2},  \\
 \label{exa}
 &=\
\frac{(3\varepsilon_{\mathrm{dd}})^{5/2}}{48}   \Re \left[(8+26\eta+33\eta^2)\sqrt{1+\eta}\right.\nonumber\\
& + \left.
\ 15\eta^3 \mathrm{ln} \left( \frac{1+\sqrt{1+\eta}}{\sqrt{\eta}}\right)  \right], \quad  \eta = \frac{1-\varepsilon_{\mathrm{dd}}}{3\varepsilon_{\mathrm{dd}}},
\end{align}
where $\Re$ denotes the real part. The dimensionless ratio 
$\varepsilon_{\mathrm{dd}}$, viz. Eq. (\ref{qf}),  
determines
the strength of  dipolar interaction relative to  that of  contact interaction 
and is useful to classify and study  many properties of a dipolar BEC. In the present study of a strongly dipolar BEC 
$\varepsilon_{\mathrm{dd}}>1 $, while  $Q_5(\varepsilon_{\mathrm{dd}})$ 
of Eqs. (\ref{exa1}) and (\ref{exa}) are complex with a small imaginary part and  a much larger real part \cite{young}.
The imaginary part contributes to a loss of atoms from the BEC  and will be neglected as in all other investigations \cite{th4,th2,th3,th5,th1}. In this study we will use Eq. (\ref{exa}) in our numerical calculation.

It is convenient to present Eq. (\ref{eq.GP3d}) in  the following  dimensionless form by scaling  lengths in terms of   $l = \sqrt{\hbar/m\omega_z}$, time in units of $t_0\equiv \omega_z^{-1}$,  angular frequency in units of $\omega_z$,  energy in units of $\hbar\omega_z$
and density $|\psi|^2$ in units of $l^{-3}$
\begin{align}\label{GP3d2}
\mbox i \frac{\partial \psi({\bf r},t)}{\partial t} & =
{\Big [}  -\frac{1}{2}\nabla^2
+{\frac{1}{2}}\left({f_x^2}x^2+ {f_y^2}y^2+z^2\right)
\nonumber \\
&+ 
4\pi{a} N \vert \psi({\bf r},t) \vert^2
+3a_{\mathrm{dd}}  N 
\int 
U_{\mathrm{dd}}({\bf R})
\vert\psi({\mathbf r'},t)\vert^2 d{\mathbf r}'   \nonumber 
\\ 
&+
\gamma_{\mathrm{LHY}}N^{3/2}
|\psi({\mathbf r},t)|^3
\Big] \psi({\bf r},t), 
\end{align} 
where all variables are  scaled. We will  use the same symbols to represent the scaled and unscaled variables  without any risk of confusion. 

The dimensionless improved mean-field GP  equation (\ref{GP3d2})  
can be derived  from   the following variational principle 
\begin{align}
\mathrm i \frac{\partial \psi}{\partial t} = \frac{\delta E}{\delta \psi^*}, 
\end{align}
where the energy functional $E$ given by 
\begin{align}
E &= \frac{1}{2}\int d{\bf r} \Big[ {|\nabla\psi({\bf r})|^2} +\Big({f_x^2}x^2+ {f_y^2}y^2+z^2\Big)|\psi({\bf r})|^2\nonumber  
 \\
&+ {3}a_{\mathrm{dd}}N|\psi({\bf r})|^2 
\left. \int U_{\mathrm{dd}}({\bf R} )
|\psi({\bf r'})|^2 d {\bf r'} \right. \nonumber \\
& + 4\pi Na |\psi({\bf r})|^4 +\frac{4\gamma_{\mathrm{LHY}}}{5} N^{3/2}
|\psi({\bf r})|^5\Big]
\end{align}
is the energy per atom of 
 a stationary state.

\label{III}   

\begin{figure}[t!]
\begin{center}

\includegraphics[width=\linewidth]{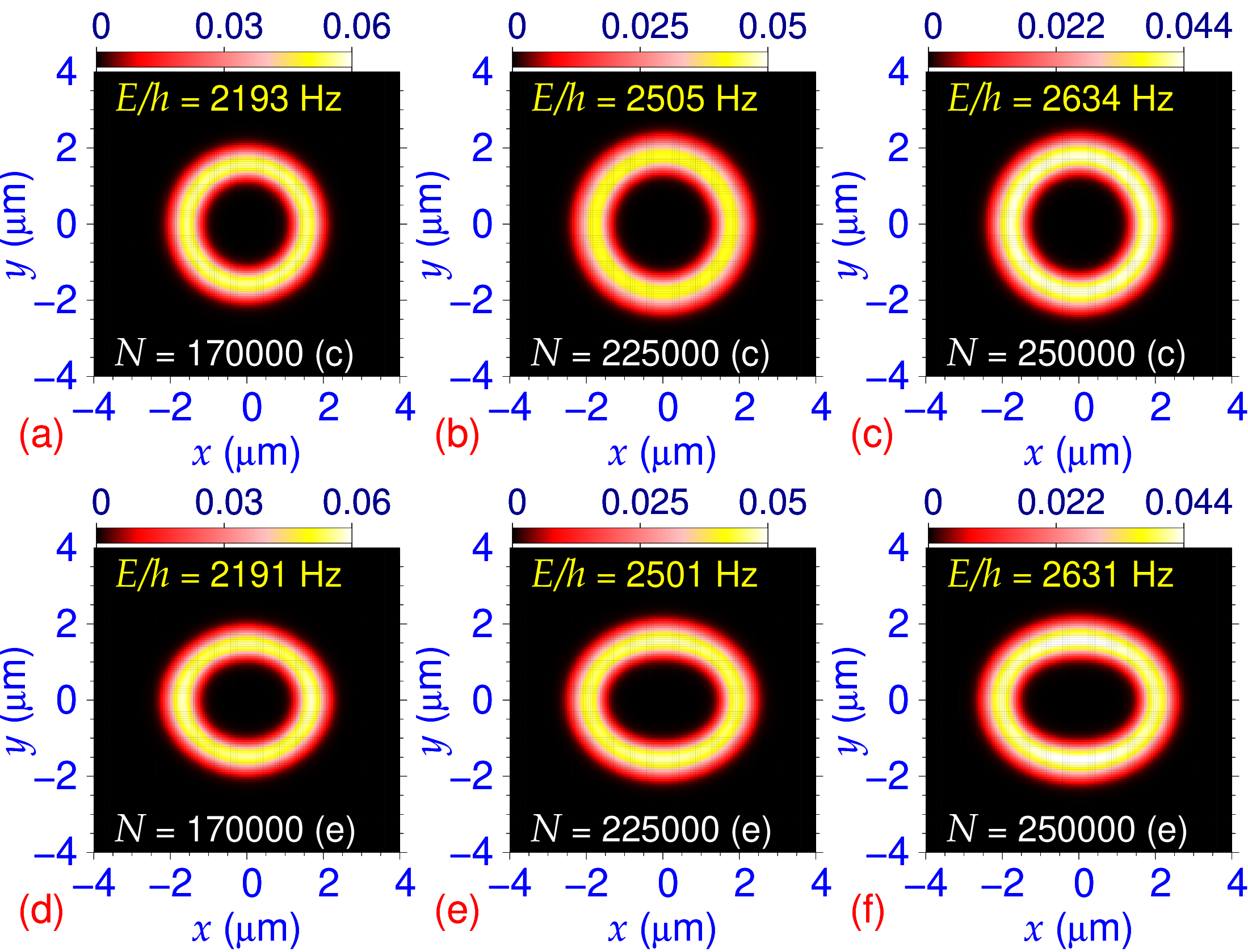}

\caption{ 
Contour plot of dimensionless 2D density $n_{\mathrm{2D}}(x, y)$  of $^{164}$Dy atoms of cylindrical-shell-shaped states for (a) and (d)  $N = 170000$,  (b) and (e)  $N =225000$,
 (c) and (f)  $N =250000$. In all cases $f_z=167$ Hz, in (a)-(c) $f_x=f_y=0.75f_z$, establishing three systems with circular (ci) section and in (d)-(f) $f_x=120$ Hz and $f_y=130$ Hz for three states with elliptic (el) section. The corresponding energies per atom $E$ (in units of $h$) are shown in the inset of each plot.
}

\label{fig1} 
\end{center}
\end{figure}

\section{Numerical Results}

\label{III}

\begin{figure*}[htbp]
\begin{center}
\includegraphics[width=\textwidth]{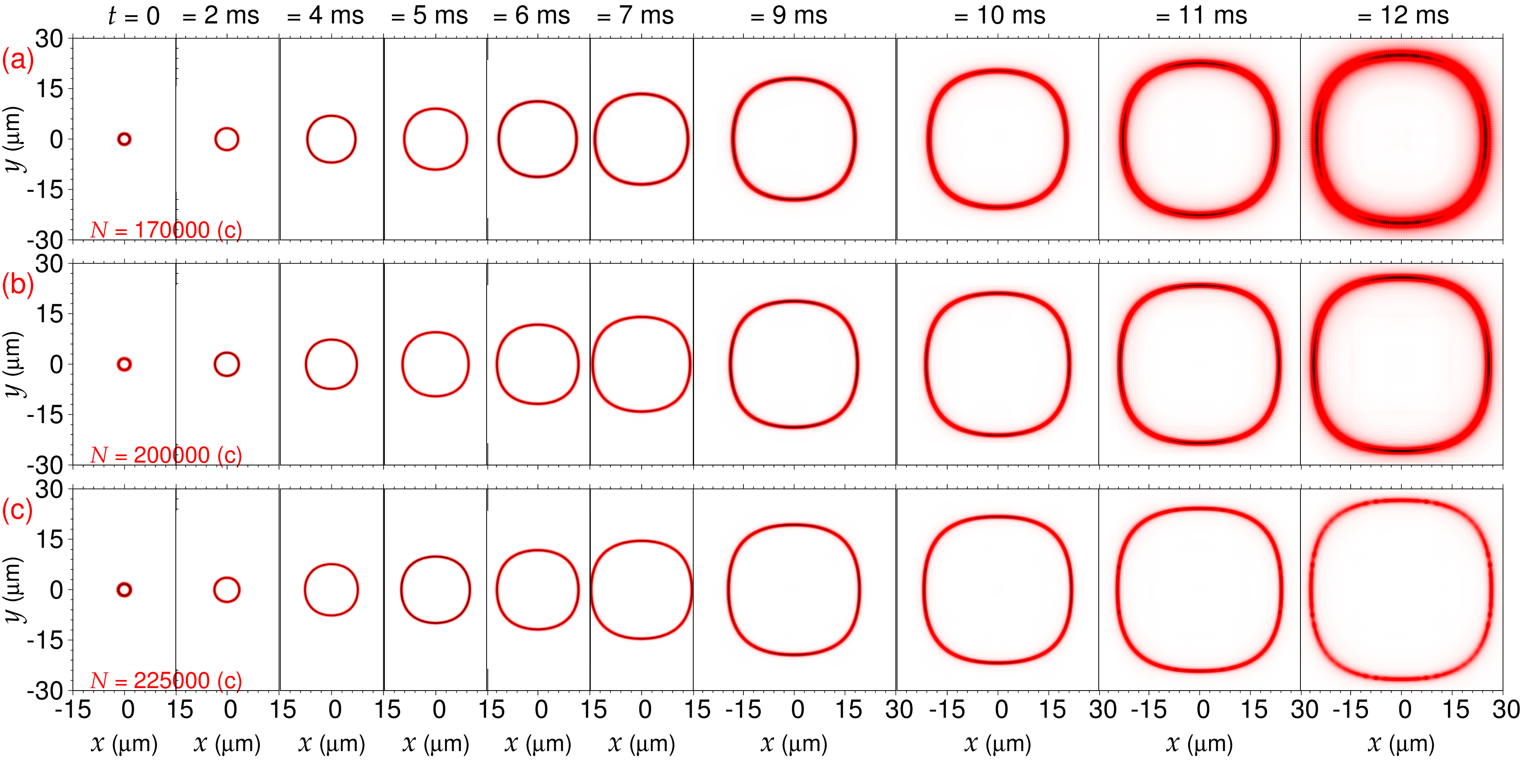}

\caption{Profile of the cylindrical-shell-shaped strongly dipolar $^{164}$Dy BEC with circular  section  during quasi-free expansion at times $t=$ 0, 2 ms, 4 ms, 5 ms, 6 ms, 7 ms,  9 ms, 10 ms, 11 ms, 12 ms,  through a contour plot of the quasi-2D density $n_{2\mathrm D}(x,y)$ for (a) $N=170000$ (top panel), (b) $N= 200000$ (middle panel), and (c) $N= 225000$ (bottom panel).  The dynamics is studied by real-time propagation, with the axial trap  $f_z=167$ Hz and radial traps $f_x=f_y=0$, using the appropriately trapped  ($f_z=167$ Hz, $f_x=f_y=0.75f_z$)   converged imaginary-time solution as the initial state at $t=0$.}

\label{fig2} 
\end{center}
\end{figure*}

To study the expansion dynamics of a cylindrical-shell-shaped 
 dipolar BEC  of $^{164}$Dy atoms,
the   partial differential  GP  
 equation (\ref{GP3d2}) is solved,  numerically, using FORTRAN or C programs \cite{dip} or specially their open-multiprocessing   counterparts \cite{ompF,omp}.  We   employ   the split-time-step Crank-Nicolson
method  by real-time propagation \cite{crank}  using the converged  wave function obtained by imaginary-time propagation as the initial state. It is numerically difficult to deal with the divergent $1/|{\bf R}|^3$ term in the dipolar potential (\ref{dip-pot}) in configuration space.  To overcome this difficulty, we evaluate the integral over the long-range dipolar potential in the improved mean-field model (\ref{GP3d2}) by a Fourier transformation to momentum space.    In this fashion, after solving the problem  in momentum space,  the configuration space solution  is obtained by  a backward Fourier transformation \cite{dip}.

 Although,  $a_{\mathrm{dd}}=130.8a_0$ for  $^{164}$Dy atoms, 
we have a certain flexibility in fixing the scattering length $a$, as the scattering length can be modified employing  
the Feshbach resonance \cite{fesh} technique by manipulating an external electromagnetic field.
In this study, as in Ref. \cite{hc},  
 we take the  scattering length $a=80a_0$. 
 The  experimental estimate of the scattering length,    $a=(92\pm 8)a_0$  \cite{scatmes}, leads to a 
 much stronger contact repulsion and  does not allow the formation of a cylindrical-shell-shaped BEC.
 With the reduction of contact repulsion,  
 the present  choice of scattering length ($a=80a_0$)   facilitates the formation of a pronounced cylindrical-shell-shaped BEC.
 
In this study of cylindrical-shell-shaped states in a strongly dipolar BEC of $^{164}$Dy atoms,  
we will consider the axial trap frequency
$f_z=167$ Hz, which is the same used in the pioneering experiments on 2D hexagonal supersolid formation with $^{164}$Dy atoms \cite{ss2,prl2d} and also used in some theoretical investigations \cite{hc,th4,th1}. 
 It is
true that the frequency $f_z$ is the same in both cases, but the quasi-2D trap of Refs. \cite{ss2,prl2d} ($f_z\gg f_x,f_y$)
is  
very different from the trap in the present study ($f_x,f_y = 0.75f_z$)  with much larger  trap frequencies along  the $x$ and $y$ directions. Cylindrical-shell-shaped states are formed for $f_z \gtrapprox 150$ Hz; however, for larger $f_z$ the inner radius of the cylindrical shell gradually reduces and eventually solid cylindrical states are formed for $f_z\gtrapprox 275$ Hz.  On the other hand, for smaller $f_z$ ($f_z \lessapprox  150$)  four droplet-states are preferentially formed. 
This is why in this study we will consider only $f_z=167$ Hz, where pronounced hollow cylindrical states with thin shell and large internal radius are formed.
For    $m(^{164}$Dy) atoms  with mass   $\approx 164 \times 1.66054\times 10^{-27}$ kg, $\hbar =  1.0545718 \times  10^{-34}$ m$^2$ kg/s, and for $f_z =  167$ Hz,   the unit of length is $l=\sqrt{\hbar/(2\pi m f_z)}= 0.6075$ $\mu$m, { the unit of time $t_0\equiv (2\pi \times f_z)^{-1} = 0.953$ ms.    }

\begin{figure*}[htbp]
\begin{center}
\includegraphics[width=\textwidth]{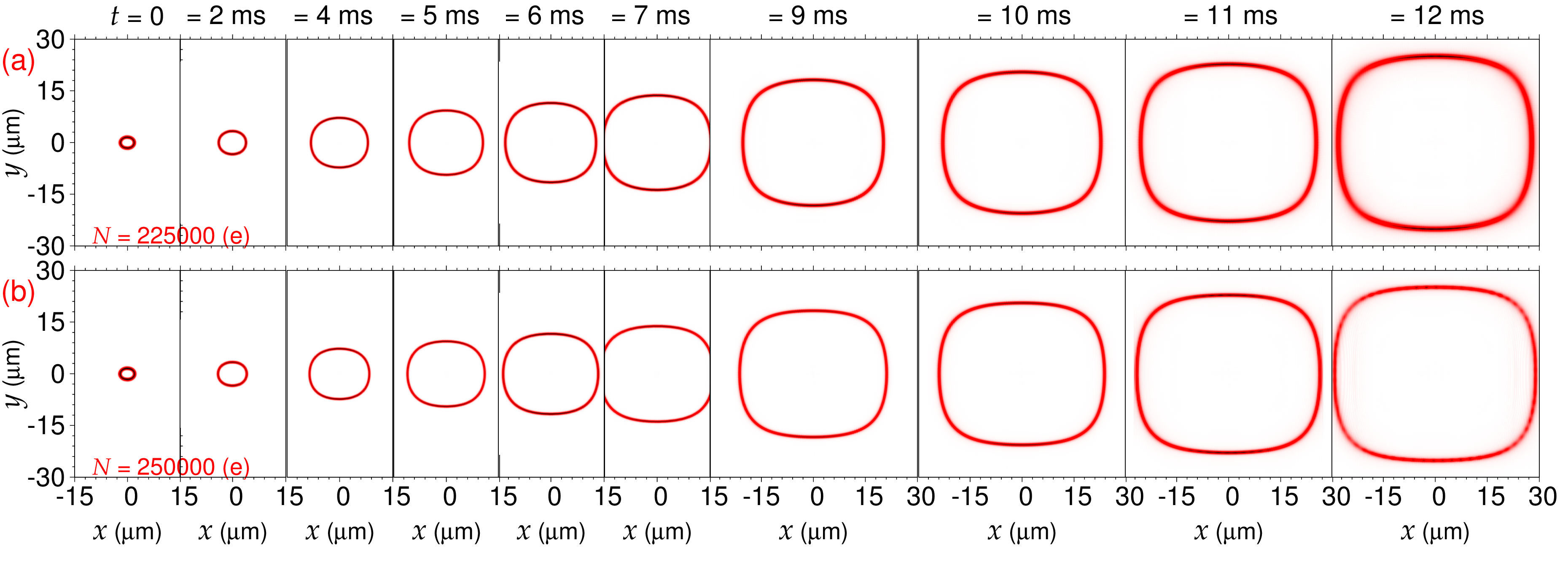}

\caption{Profile of the cylindrical-shell-shaped strongly dipolar $^{164}$Dy BEC with elliptic  section  during quasi-free expansion at times $t=$ 0, 2 ms, 4 ms, 5 ms, 6 ms, 7 ms,  9 ms, 10 ms, 11 ms, 12 ms,  through a contour plot of the quasi-2D density $n_{2\mathrm D}(x,y)$ for (a)  $N=225000$ (top panel) and  (b) $N=250000$  (bottom panel).  The dynamics is studied by real-time propagation, with the axial trap  $f_z=167$ Hz and the radial traps $f_x=f_y=0$, using the appropriately trapped  ($f_x= 120$ Hz, $f_y=130$ Hz)  converged imaginary-time solution as the initial state at $t=0$.}

\label{fig3} 
\end{center}
\end{figure*}

A  cylindrical-shell-shaped 
state  is best  illustrated through  the  integrated 2D density $n_{{\mathrm{2D}}}(x,y)$ defined by an axial $z$ integration
over density
\begin{align}
n_{{\mathrm{2D}}}(x,y)=\int_{-\infty}^{\infty} dz |\psi(x,y,z)|^2.
\end{align} 
 A contour plot of the 2D density $n_{{\mathrm{2D}}}(x,y)$   of a few   cylindrical-shell-shaped states with circular (ci) and elliptic (el) sections  in a dipolar BEC of $^{164}$Dy atoms  is considered next for $a=80a_0, f_z=167$ Hz.
 In  Fig. \ref{fig1}, we display  a contour plot of the 2D density $n_{{\mathrm{2D}}}(x,y)$ of 
  a cylindrical-shell-shaped state with circular section for  (a)   $N=170000,$ (b) $N=200000$ and (c) 
 $N=225000$   and $f_x=f_y=0.75f_z$. Next we display the same  with elliptic section  for   (d)   $N=170000,$ (e) $N=225000$ and (f) 
 $N=250000$   and $f_x=120$  Hz and $f_y=130$ Hz.  The trap in plots \ref{fig1}(a)-(c) is cylindrically symmetric whereas this symmetry of the trap is broken in  plots \ref{fig1}(d)-(f). Consequently, 
 we find cylindrical-shell-shaped states in   plots \ref{fig1}(a)-(c) with circular section   and in plots  \ref{fig1}(d)-(f) with elliptic section. In all cases 
 the inner radius of the cylindrical shell is quite sharp and pronounced.    The corresponding energies, given in the insets of the plots, of the elliptical cylinders are slightly smaller than those of the circular cylinders.

  In a typical experiment, the cylindrical-shell-shaped state would be confirmed  after an expansion to a detectable size  by  a two-step process. 
  We study by real-time  propagation a quasi-free expansion in the $x$-$y$ plane, by making $f_x=f_y=0$ at $t=0$, while we maintain the BEC trapped 
along the $z$ direction.     As the cylindrical shape appears in the $x$-$y$ plane, such a quasi-free expansion in this plane is appropriate for our purpose. After a quasi-free expansion in the $x$-$y$ plane, the frequency of the trap in the $z$ direction should be reduced/relaxed  a bit for the cylindrical-shell-shaped state to expand in the $z$ direction up to a desired size to visualize (not elaborated in this paper).  In a fully free expansion ($f_x=f_y=f_z=0$) from the beginning,
the  cylindrical-shell-shaped BEC expands more in the $z$ direction.  As the intention is to have a large curved surface after expansion, this two-step process is more appropriate.

The quasi-free expansion dynamics as obtained by real-time propagation is presented in Fig. \ref{fig2} for the cylindrical-shell-shaped states with circular section for (a) $N=170000$, (b)  $N=200000,$ and (c) $N=225000$ $^{164}$Dy atoms through a contour plot of  the 2D density $n_{{\mathrm{2D}}}(x,y)$ at times  $t=0,$ 2 ms, 4 ms, 5 ms,
6 ms, 7 ms, 9 ms, 10 ms,  11 ms, and 12 ms. The typical diameter of the cylindrical-shell-shaped state has increased from about 4 $\mu$m to 60 $\mu$m in 12 ms which will guaranty the observability of these states.  The thickness of the cylindrical shell has increased a bit for $N=170000$, although the internal hollow region is very prominent. With the increase of the number of atoms, after expansion, the cylindrical shell 
remains thin and sharp, viz. the bottom panel in Fig. \ref{fig2} for $N=225000$.

  Next we study the quasi-free expansion of a cylindrical-shell-shaped state with elliptic  section. We display in Fig. \ref{fig3} this expansion dynamics  of (a) $N=225000$ and (b) $N=250000$ $^{164}$Dy atoms  through a contour plot of  the 2D density $n_{{\mathrm{2D}}}(x,y)$ at times  $t=0,$ 2 ms, 4 ms, 5 ms,
6 ms, 7 ms, 9 ms, 10 ms,  11 ms, and 12 ms. The sharpness of the shell structure after expansion in both cases in Fig. \ref{fig3} are quite similar and is very pronounced. The expansion of the cylindrical-shell-shaped states in Fig. \ref{fig3} is comparable to the same in Fig. \ref{fig2}.    The pronounced shell structure in Figs. \ref{fig2} and \ref{fig3} are due to the robustness of these states with strong dipolar interaction. The dipolar repulsion between the $^{164}$Dy   atoms in the same $x$-$y$ plane and the  dipolar attraction between the atoms in different  $x$-$y$ planes make the robust cylindrical-shell-shaped states. The same will not be possible in the absence of a dipolar interaction. This is why the expansion of a nondipolar repulsive
spherical-shell-shaped  BEC \cite{expan2}  or a ring-shaped BEC \cite{expan5}  may easily  destroy the  internal hollow region.

\begin{figure}[t!]
\begin{center}

\includegraphics[width=\linewidth]{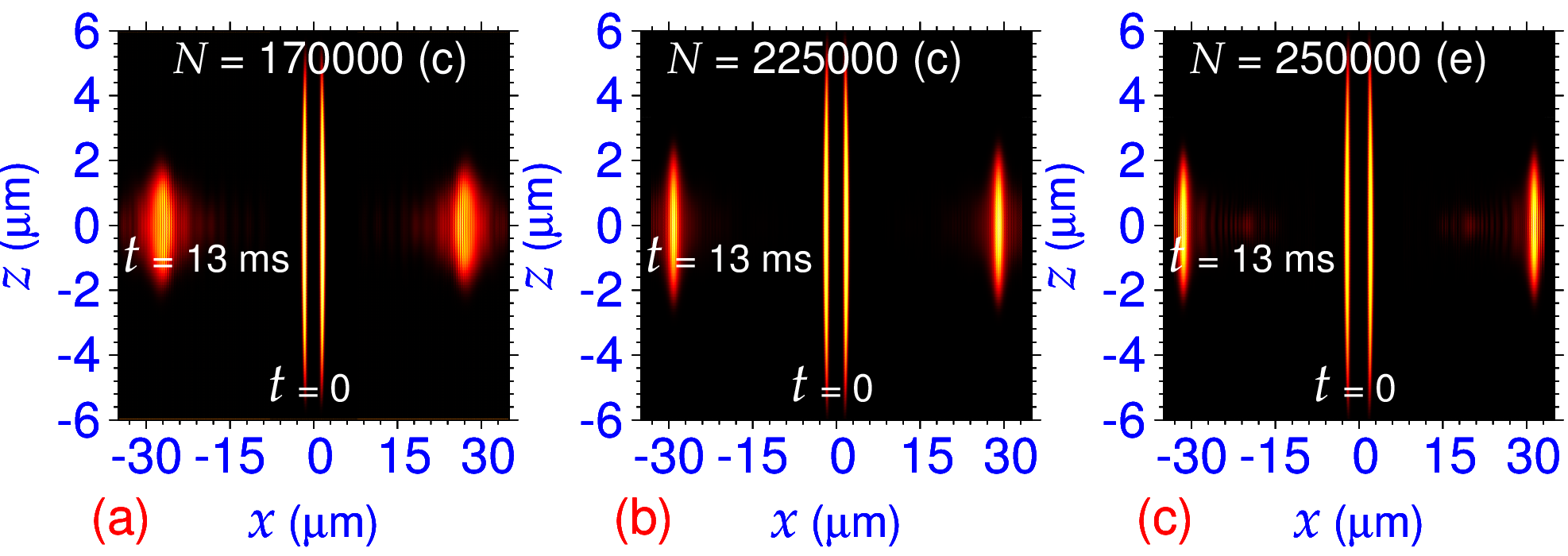} 

\caption{Contour plot of density $|\psi(x,0,z)|^2$ of the quasi-freely expanding cylindrical-shell-shaped dipolar BEC of  $^{164}$Dy atoms  at times $t=0$ (central region with small $|x|$) and $t=13$ ms  (outer region with large $|x|$)    
for (a) $N=170000$, (b) $N=225000$, and (c) $N=250000$. In (a) and (b) the section is circular (ci) corresponding to the expansion dynamics of Figs. \ref{fig2}(a) and (c),   and in (c) the section is elliptic (el) corresponding to the expansion dynamics of Fig.  \ref{fig3}(c). 
}
\label{fig4} 
\end{center} 
\end{figure}

\begin{figure}[t!]
\begin{center}

\includegraphics[width=.9\linewidth]{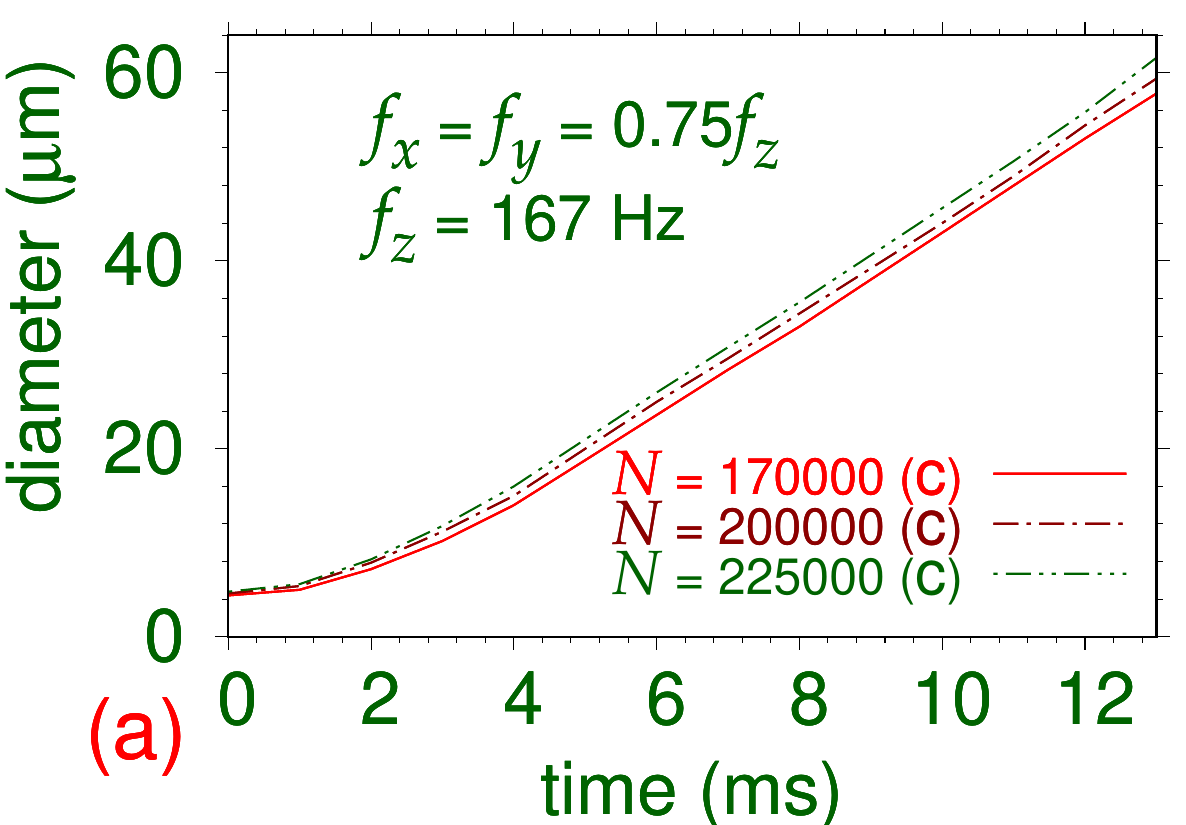}
\includegraphics[width=.9\linewidth]{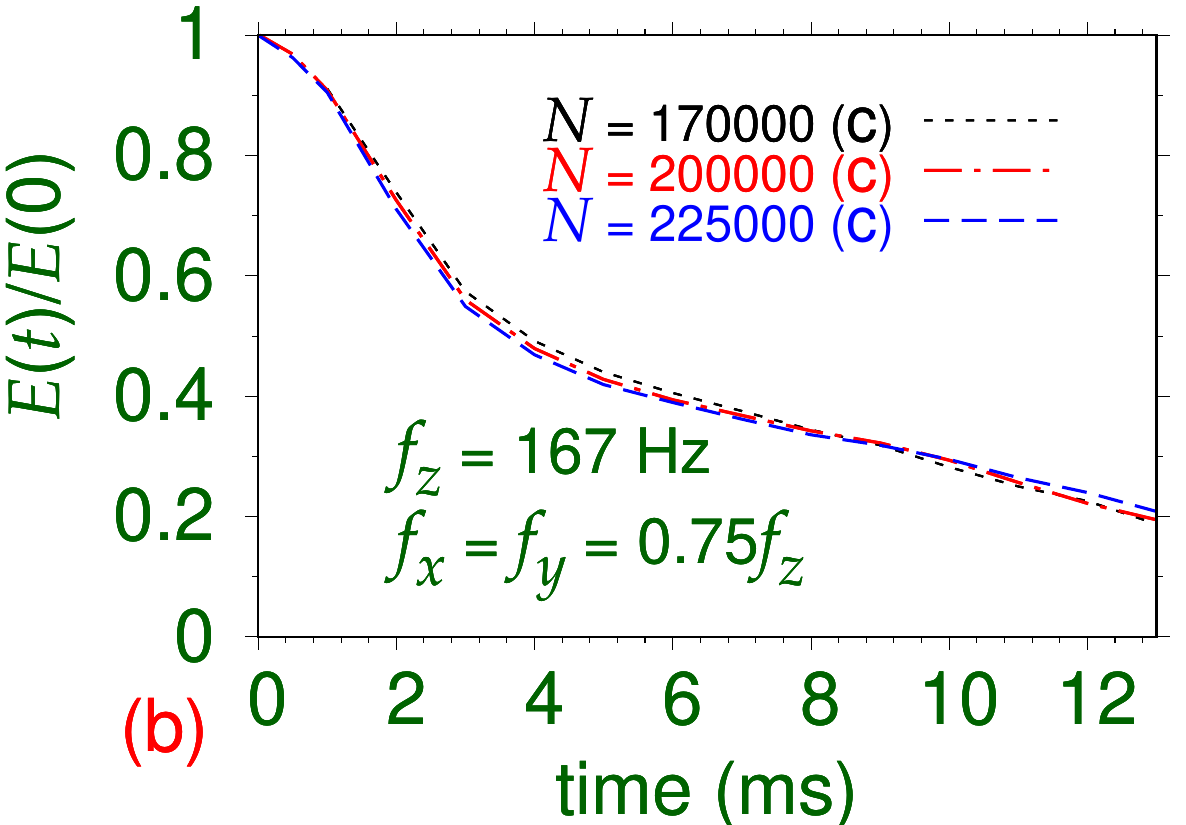}

\caption{(a)  Evolution of the outer diameter  in the $x$-$y$ plane 
of the  cylindrical-shell-shaped dipolar BEC  of $^{164}$Dy atoms with circular section during a quasi-free expansion for different $N=170000, 200000,$ and 225000. (b) Evolution of the energy $E(t)/E(0)$ of the same  
 during this expansion in units of respective energies at $t=0$ $E(0)$:  for $N=170000,$ $E(0)/h=2193$ Hz;
  for $N=200000$, $E(0)/h=2368$ Hz;  and   for $N=225000$, $E(0)/h=2505$ Hz.
} 
\label{fig5} 
\end{center} 
\end{figure}

 The extended hollow region of the cylindrical-shell-shaped state is best illustrated through a 
 contour plot of density $|\psi(x,0,z)|^2$  of the quasi-freely expanding cylindrical-shell-shaped dipolar BEC of $^{164}$Dy  atoms at $t=0$ and at $t=13$ ms. This is displayed 
 in Fig. \ref{fig4} for (a) $N=170000$, (b) $N=225000$  of  a cylindrical-shell-shaped BEC of circular section, viz.  Figs. \ref{fig2}(a) and (c),
 and in Fig. \ref{fig4} for (c) $N=250000$ of a cylindrical-shell-shaped BEC of elliptic section, viz. Fig. \ref{fig3}(b). In all these plots the long and thin $y=0$ section of the cylinder at $t=0$ can be seen in the central region of small $|x|$ values, whereas in the outer large $|x|$ region, the short and wide section of the cylinder at $t=13$ ms  can be seen.  For larger $N$ in Figs. \ref{fig4}(b) and (c), for $N=225000$ and $N=250000$, the hollow cylinder remains sharp with  thin wall, whereas the same is a bit blurred in Fig. \ref{fig4}(a), for $N=170000$, with wider wall.  This could have been anticipated from the  $t=12$ ms profile of the same state in Fig. \ref{fig2}(a).
  In all cases, after expansion, as the diameter of the cylinder increases, its length has reduced. 
 A study of the diameter and energy of the expanding cylindrical-shell-shaped  BEC of $^{164}$Dy atoms of circular section  is considered next.  In Fig. \ref{fig5}(a)   we plot the outer diameter  of the expanding hollow cylindrical states of  circular section  displayed in Fig. \ref{fig2} versus time $t$. In Fig. \ref{fig5}(b) we plot the energy (per atom) $E(t)$  of the same states in unit of the initial $t=0$ energy $E(0)$ versus $t$. In Fig. \ref{fig5}(a) the diameter increases monotonically with time,   whereas  the energy  decreases monotonically with time in Fig. \ref{fig5}(b) during expansion.   At $t=13$ ms the diameter (size) of the expanding state is about 60 $\mu$m, large enough for its experimental confirmation. 
 In Fig. \ref{fig5}(b) the evolutions of normalized energy  $E(t)/E(0)$ for $N=170000, 200000,$ and 225000 are quite similar although the initial energy $E(0)/h$ of these  states are very different.

\begin{figure}[t!]
\begin{center}

\includegraphics[width=.9\linewidth]{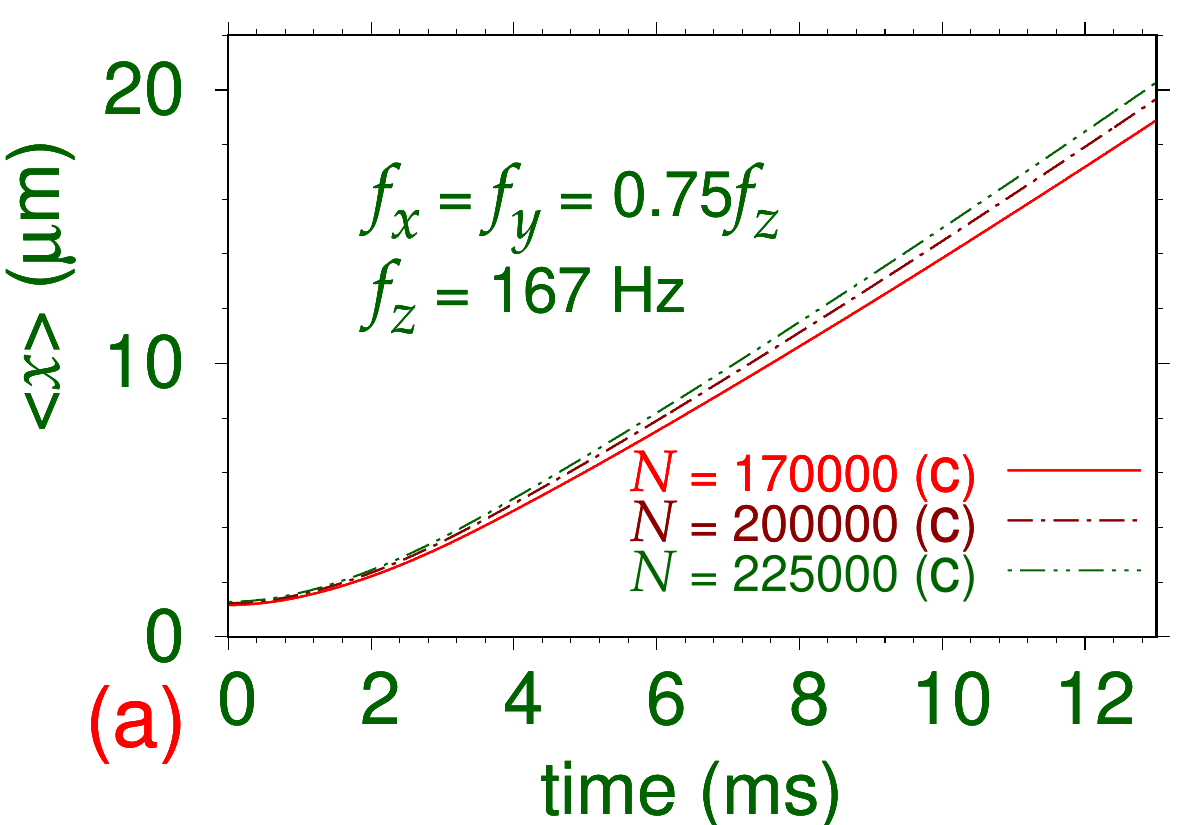}
\includegraphics[width=.9\linewidth]{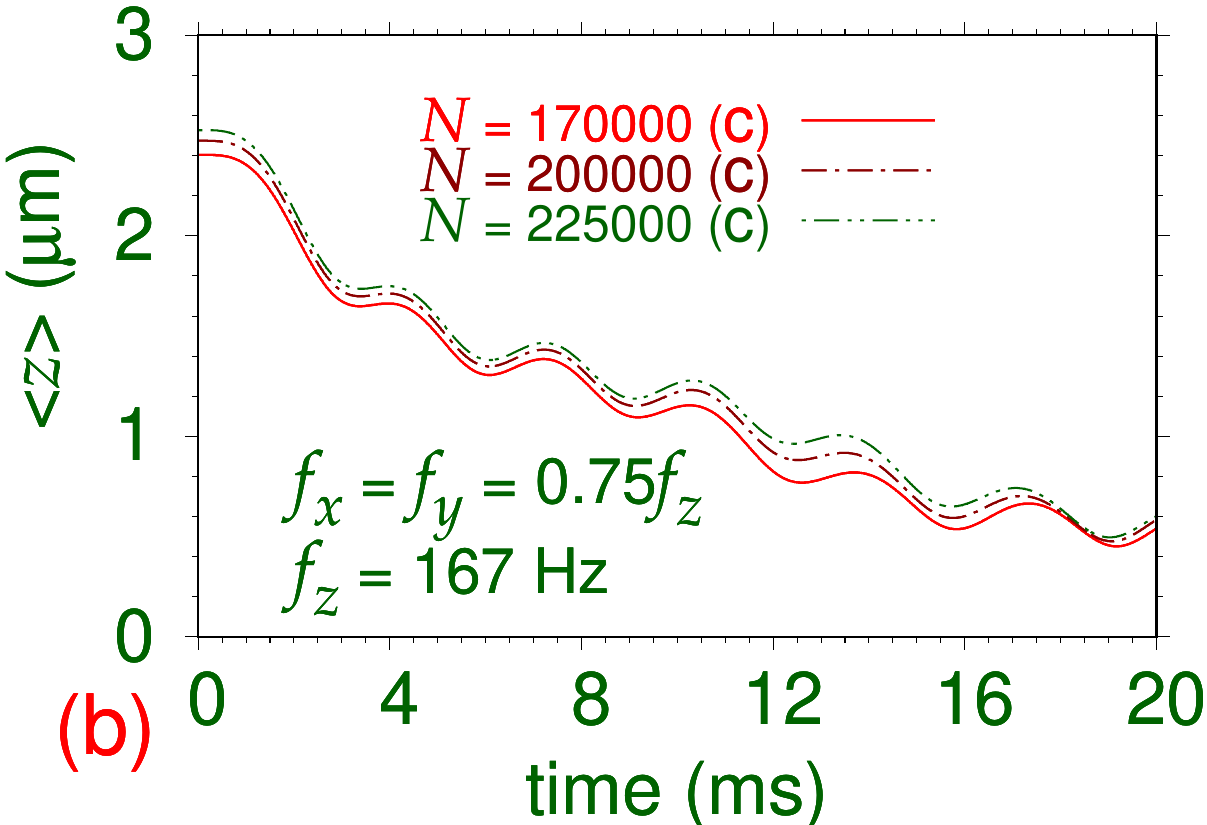}

\caption{  Evolution of the rms  size  (a) $\langle x \rangle$ and (b) $\langle z \rangle$    
of the  cylindrical-shell-shaped dipolar BEC  of $^{164}$Dy atoms with circular   section during the quasi-free expansion 
displayed in Fig. \ref{fig2}
for different $N=170000, 200000,$ and 225000.
}
\label{fig6} 
\end{center} 
\end{figure}

 Finally, we present in Fig. \ref{fig6} the evolution of the rms size   (a) $\langle x \rangle$ and 
(b)  $\langle z \rangle$ during the quasi-free expansion of the cylindrical-shell-shaped  BEC of $^{164}$Dy atoms of circular section. Not quite unexpected,  the rms size $\langle x \rangle$  has the same behavior as the diameter, viz. Fig. \ref{fig5}(a). In Fig. \ref{fig6}(b) we find that the rms size  $\langle z \rangle$ decreases during expansion but with a periodic oscillation with period $T_{\mathrm{numeric}}\approx 3.25$ ms. This is because the present expansion is only quasi-free with a trap of frequency $f_z=167$ Hz in the $z$ direction. Consequently, the expansion dynamics results in a periodic modulation in  $\langle z \rangle$ \cite{stringari}.
 This period can be compared with the theoretical period of oscillation of a cylindrically symmetric ($f_x=f_y$) trapped   repulsive nondipolar BEC  in the hydrodynamic regime \cite{stringari}
 \begin{eqnarray}
 f^2=  f_x^2  \left(2+\frac{3}{2}\lambda^2 \mp \frac{1}{2} \sqrt{9\lambda^4 -16\lambda^2+16} \right)
 \end{eqnarray}
 where $\lambda=f_z/f_x$. In the present case of extreme disk-type geometry during the quasi-free expansion $f_x\to 0$ and $\lambda \to \infty$, the only surviving frequency is $\sqrt 3 f_z$, which corresponds, for $f_z= 167$ Hz, used in this study, to the period $T_{\mathrm{theory}}=1/ (\sqrt 3 f_z)\approx  3.54$ ms, in qualitative agreement with the period of $T_{\mathrm{numeric}}\approx  3.25$ ms found in Fig. \ref{fig6}(b).  This agreement is satisfactory considering the fact that the analysis of Ref. \cite{stringari} considered a repulsive BEC with contact interaction only, whereas the present study has a sizable amount of dipolar attraction and the LHY interaction.

{   Figures \ref{fig5} and \ref{fig6} presents a comprehensive illustration of the expansion dynamics, which is 
practically insensitive to a variation of the number of atoms.  The variation of other parameters $-$ the trapping frequency $f_z$ and the scattering length $a$ $-$ are not of much interest from a phenological point of view.   If the frequency $f_z$ is reduced the cylindrical-shell-shaped structures disappear; if it is increased the internal radius of the cylinder reduces and for a large enough $f_z$ the solution is a solid cylinder \cite{hc}. 
We established \cite{hc}   that for the formation of pronounced shell-shaped solutions  one should have 170 Hz $>f_z>140$ Hz
and $f_x\approx f_y = 0.75 f_z$.
The scattering length of the dipolar atoms cannot also be varied arbitrarily; we used the value $a=80a_0$.   If the scattering length is reduced a bit to $a=70a_0$, the delicate balance between the dipolar attraction and the contact repulsion will be lost and the system will be much too attrctive and will collapse.    If it is increased a bit to $a=90a_0$ the solution has a Gaussian shape. Hence a reasonable variation of  the frequency $f_z$ or of the scattering length $a$ will destroy the cylindrical-shell-shaped solutions.
 }

\section{Summary}

  
  In this paper we have studied the expansion dynamics of a high-density dynamically-stable  cylindrical-shell-shaped
strongly dipolar BEC of  $^{164}$Dy atoms  \cite{hc}
 for parameters $-$ number of atoms, trap frequencies $-$ quite similar to those employed in  recent experiments \cite{ss2,expt}. We  employed  an improved mean-field GP model including the LHY interaction \cite{lhy} appropriately modified for dipolar atoms \cite{qf1,qf2}.  We solved the model partial differential equation  (\ref{GP3d2}) by real-time propagation 
and suggest   a   two-step procedure to carry out the expansion in experiment.
 We considered a quasi-free expansion of the system radially in the $x$-$y$ plane setting the radial trap frequencies 
 $f_x=f_y=0$ and maintaining the axial trap frequency  $f_z$ intact.   This procedure will allow a radial expansion of the cylindrical-shell-shaped BEC up to a desirable size.     The cylindrical shell is found to expand, practically without deformation, over a large period of time.      After an expansion during 13 ms, the diameter of the cylindrical shell expands from about 4 $\mu$m to 60 $\mu$m maintaining the initial symmetry of the cylindrical-shell-shaped state with 
 both circular and elliptic sections.  The elliptic symmetry was obtained by introducing a small anisotropy of the trapping potential in  the $x$-$y$ plane.   In all cases studied,  as the cylindrical-shell-shaped state expands radially, its axial length contracts a little. During the quasi-free expansion dynamics, the rms size in the polarization $z$ direction is found to undergo a periodic oscillation controlled by the frequency $f_z$ 
of the trap in the $z$ direction, in accord with a theoretical study \cite{stringari}. 
Finally, the $z$ frequency of the trap should be reduced slightly to allow the cylinder to expand axially due to dipolar interaction so as to attain the desired size.
  The steady quasi-free expansion of the cylindrical-shell-shaped
strongly dipolar BEC of  $^{164}$Dy atoms  over a long period of time     ensures the possibility of observing  such a  state in a laboratory 
 in the near future.

\section*{CRediT authorship contribution statement}

L. E. Young-S.: Methodology, Validation, Investigation, Writing – review and  editing,
Visualization.

S. K. Adhikari: Conceptualization, Methodology, Validation, Investigation, Writing – original draft, Writing – review and  editing, Supervision, Funding aquisition, 
Visualization.

\section*{Declaration of competing interest}
The authors declare that they have no known competing financial interests or personal relationships that could have appeared
to influence the work reported in this paper.

\section*{Data availability}
No data was used for the research described in the article.

\section*{Acknowledgments}
LEY-S. would like to acknowledge the financial support by the Vicerrectoría de Investigaciones - Universidad de Cartagena through Project No. 004-2023.
SKA
 acknowledges support by the CNPq (Brazil) grant 301324/2019-0. The use of the supercomputing
cluster of the Universidad de Cartagena, Cartagena, Colombia  is gratefully acknowledged.


\end{document}